# How Deep Are the Fakes?
# Focusing on Audio Deepfake: A Survey

ZAHRA KHANJANI, GABRIELLE WATSON, and VANDANA P. JANEJA, University of Maryland Baltimore County, Information System department, USA

Deepfake is content or material that is synthetically generated or manipulated using artificial intelligence (AI) methods, to be passed off as real and can include audio, video, image, and text synthesis. This survey has been conducted with a different perspective compared to existing survey papers, that mostly focus on just video and image deepfakes. This survey not only evaluates generation and detection methods in the different deepfake categories, but mainly focuses on audio deepfakes that are overlooked in most of the existing surveys. This paper's most important contribution is to critically analyze and provide a unique source of audio deepfake research, mostly ranging from 2016 to 2020. To the best of our knowledge, this is the first survey focusing on audio deepfakes in English. This survey provides readers with a summary of 1) different deepfake categories 2) how they could be created and detected 3) the most recent trends in this domain and shortcomings in detection methods 4) audio deepfakes, how they are created and detected in more detail which is the main focus of this paper. We found that Generative Adversarial Networks(GAN), Convolutional Neural Networks (CNN), and Deep Neural Networks (DNN) are common ways of creating and detecting deepfakes. In our evaluation of over 140 methods we found that the majority of the focus is on video deepfakes and in particular in the generation of video deepfakes. We found that for text deepfakes there are more generation methods but very few robust methods for detection, including fake news detection, which has become a controversial area of research because of the potential of heavy overlaps with human generation of fake content. This paper is an abbreviated version of the full survey and reveals a clear need to research audio deepfakes and particularly detection of audio deepfakes.

## 1 INTRODUCTION

Deepfakes have started impacting society in various ways. For example, deepfakes can be used to commit fraud. A CEO of a U.K energy based firm who thought he was on the phone with his German boss had asked him to send a transfer of 220,000 Euros to a Hungarian supplier [64]. The criminals had used AI based software to impersonate him, and received one transfer of money. This may clarify the importance of deepfakes, and why the attention to it has increased exponentially. A huge increase at the number of articles regarding deepfake happened between 2018 and 2019 (from 60 to 309). On 24th July, It was linearly estimated that the number of papers related to deepfakes will increase to more than 730 until the end of 2020 [51]. However, the reality is more surprising than the mentioned estimate since we found there are 1323 papers related to deepfakes that were published until the end of 2020. The lack of focus on audio deepfakes in surveys is a strong motivation for this article to concentrate on audio deepfakes, where it is heading and how to weaken its harmful effects.

The rest of the paper is organized as follows: In Section 2 we present a systematic review on the scientific papers for each category of deepfakes, their generation and detection techniques and the most recent trends. More details, the network schematics, as well as a quick guide table 1 of audio deepfake frameworks are provided in this section. Section 3 includes discussion and future directions. The summarization tables 2 and 3 of some of the significant papers that are surveyed in this work are also presented in Section 3. Finally, our conclusions are presented in Section 4. We categorize deepfakes as shown in figure 1. Then, we describe each of these types of deepfakes and ways by which they can be detected and created.

Authors' address: Zahra Khanjani, zkhanja1@umbc.edu; Gabrielle Watson, watson7@umbc.edu; Vandana P. Janeja, vjaneja@umbc.edu, University of Maryland Baltimore County, Information System department, 1000, Hilltop Circle, Baltimore, Maryland, USA, 21250.

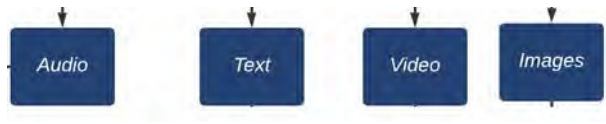

Fig. 1. Deepfake categories

## 2 DEEPFAKE CATEGORIES

For each category (audio, text, video, or image deepfakes), related papers are surveyed and the technology trends and frameworks are briefly discussed and shown in the figures. To the best of our knowledge, this paper is the first survey focusing on audio deepfakes. In the audio deepfake section we discuss some important frameworks in detail and provide readers with sufficient guidance for audio deepfake tools as are shown in Table 1. In the following sections, we begin the explanations of the deepfake categories.

### 2.1 Audio deepfakes

Audio deepfakes are AI generated or edited/synthesized to create fake audio that seems real. Detecting audio deepfakes is really important since there have been some criminal activities using audio deepfakes in recent years. To achieve audio deepfake detection, one firstly needs to know the generation methods. Figure 2 (a) shows deepfake generation frameworks and trends, while Figure 2 (b) shows audio detection tools and trends. As figure 2 indicates, audio deepfake methods are divided into three subcategories: Replay attack, speech synthesis, voice conversion. The reader is provided with the most recent and significant frameworks of each subcategory in this section.

*2.1.1 Replay attacks.* Replay attacks are defined as replaying the recording of a target speaker's voice. The two subtypes are far field detection and cut and paste detection attacks [56]. In far field detection replay attacks, the test segment is a far field microphone recording of the victim that has been replayed on a phone handset with a loudspeaker. Cut and paste detection system is if a recording is made by cut and paste short recordings to fake the sentence required by a text dependent system [56]. To defend against replay attacks one can use text dependent speaker verification [72]. A current technique that detects end-to-end replay attacks is by using deep convolutional networks [70].

Some of the replay attack detection systems have been proposed by working on the features which are fed into the network [75]. Others have improved the networks used or have worked on both of the networks and features [21, 23, 24, 38, 39, 42, 48]. Machine learning is not very effective for finding replay attacks because of overfitting due to the variability in speech signals [43]. It was found in the technique to detect replay attacks with deep convolutional networks that they were able to get a perfect Equal Error Rate(EER) of 0 percent for the development and evaluation set for ASVspoof2017 [34]. It means the performance of the detection technique was really better than the previous ones since before [70].

*2.1.2 Speech synthesis (SS).* Speech synthesis is defined as artificially producing human speech by means of software or hardware system programs. SS includes text to speech (TTS) which analyzes the text and makes the speech sound in line with text inputted using the rules of linguistic description of the text. Speech synthesis can be used for purposes like reading text and being a personal AI assistant. Another benefit is speech synthesis can offer different accents and voices instead of pre-recorded human voices. One of the leading speech synthesis companies is Lyrebird which uses deep learning models to generate 1,000 sentences in a second. TTS heavily depends on the speech corpus quality to make the system and, unfortunately, it is expensive to create speech corpora ([36]. Another disadvantage is that SS systems do not recognize periods or special characters [36]. Ambiguities with homographs are the largest, which is when two words have different meanings but are written the same way [36]. ([5]).

Char2Wav is an end-to-end speech synthesis generation framework. Also, WaveNet [52], a SS framework, is based on PixleCNN. Text-to-speech synthesis frameworks often include two phases (encoder and decoder), WaveGlow focuses on the second phase. Therefore, WaveGlow is regarding transforming some time-aligned features, such as a mel-spectrogram obtained from encoder, into audio samples [57]. Tacotron, was originally suggested in 2017 [74]. Tacotron includes CBHG which is (1-D convolution bank + highway network + bidirectional GRU) [40]. Tacotron 2 [60] system includes two components. The first component is a recurrent sequence-to-sequence feature prediction network with attention. The output of this component is a predicted sequence of mel spectrogram frames. The second component is a modified WaveNet vocoder. Conv: convolutional, Attention: Location Sensitive Attention.

Deep Voice3, a TTS framework, [55] includes 3 parts:
- Encoder: residual convolutional layers are used to encode text into per-timestep key and value vectors.
- Decoder: (key, value) is used by decoder to predict the mel-scale log magnitude spectrograms. It contains causal convolutional blocks. Mel-band log-magnitude spectrogram is used as the compact low-dimensional audio frame representation. Two loss functions are used : L1 loss based on the output mel-spectrograms, and a binary cross-entropy loss based on the final-frame prediction.
- Converter: A fully-convolutional post-processing network. Based on the chosen vocoder and the decoder hidden states, it predicts the vocoder's parameters. Conv= convolutional, dotted arrows = the autoregressive process during inference.

MelNet, a TTS framework [71], works in an autoregressive manner, and predicts a distribution element-by-element over the time and frequency dimensions of a spectrogram. The network includes different computational stacks that extract features from different pieces of the input. Then, these features will be collectively summarized to make the full context. The previous-layer's outputs of the frequency-delayed stack are current-layer's output of the time-delayed and centralized stacks respectively. The outputs of the final layer of the frequency-delayed stack are used to compute the needed parameters for the audio generation.

Using neural network TTS synthesis can make the speech audio in the voice of many speakers even those not in the training. This only needed five seconds [28]. The first model to synthesize audio directly from text was Char2Wav which is end-to-end speech synthesis which has a reader and a neural vocoder to accomplish this [62]. Deep Voice 1 was the first to operate in real time for deep neural networks for text to speech, which is the foundation for end to end neural speech synthesis[7] and Deep Voice 2 [20], was able to reproduce many voices using the same system. Moreover, most neural network based models for speech synthesis are auto regressive, meaning that they condition the audio samples on previous samples for long term modeling and are simple to train and implement [57].The SS detection systems are also used for voice conversion (VC) detection, so we review the detection methods for these two categories in the VC summary section.

*2.1.3 Voice conversion (VC) and Impersonation.* The last subcategory of audio deepfakes is voice conversion, which takes the speech signal by the first speaker, the source, and modifies it to sound like it was spoken by the second speaker, i.e., the target speaker. Impersonation that can be considered as a kind of VC is pretending to be another person. it is faster now to impersonate with new technology and one company called Overdub[5] can do an impression of any voice with one minute of sample audio. Also, GANs can be used for voice impersonation [19].

One of the most important bases for VC is the joint density Gaussian mixture model with maximum likelihood parameter trajectory generation considering global variance [67]. This model is also the baseline of the open-source Festvox system that was the main VC toolkit in " The voice conversion challenge 2016" [68]. Voice conversion can be also based on other methods such as neural networks as well as speaker interpolation [27, 49, 67]. However, in recent years GANs are widely used for VC due to their flexibility as well as high-quality results. The research of [19] used a neural network framework to impersonate voices from different genders well with reconstructing time domain signals with the Griffin Lim method. This led to the model creating very convincing samples of impersonated speech. There are also different frameworks for audio spoofing detection. ResNet which was firstly used for image recognition, is utilized as the base of the audio spoofing (VC and SS) detection system [13]. It is also improved to reduce EER metric as well as solve the generalization problem [12]. In addition, some of the audio spoofing detection systems have been extended by working on the features which are fed into the network [8]. While others have worked on the networks used or both of the networks and features [6, 14, 44, 53, 59, 73].

Quick guide table 1 is provided, and it contains different audio deepfake tools, their summarized key features as well as high-starred GitHub repository links.

## 2.2 Text Deepfake

The text deepfake field is teeming with papers and techniques to create deepfakes, and detection methods are catching up but not fast enough.

**Exposed fabrications** One of the subcategories of a textual deepfake is exposed fabrications which are those that are being fraudulently reported, like tabloids and yellow press which use sensationalism and eye-catching headlines to get more profit/traffic utilizing AI methods.

**Humorous fakes** The next subcategory of textual deepfakes are humorous fakes which is information that if the readers do not know that it is of humorous intent, they might take the information at face value.

**Large hoax** The last subcategory of textual deepfakes is the large hoax which is falsification or deliberate fabrication in mainstream media that attempts to deceive audiences that it is real news which can be picked up by traditional news outlets. Figure 3 presents a summary of different text deepfake generation and detection frameworks and features.

## 2.3 Video deepfake

A summary for video deepfakes generation and detection frameworks and features are provided in Figure 4. Generally, video editing has been around for a while since 1997 in Forrest Gump to digitally put in archival footage of JFK and manipulate his mouth movements[4]. Later, deepfake technology made video editing more believable. Many YouTube channels like Ctr Shift Face [2] post video deepfakes with increasing capabilities due to the new tools and the vast amount of training data that is available on the internet.

**Reenactment** The first subcategory of video deepfakes is reenactment in which a person manipulates the identity to impersonate it and control what the identity says or does for the expression, body, dubbing, pose, and gaze. The expression refers to the reenactment which drives the expression of the face. The mouth reenactment is also called 'dubbing' or lip sync. The pose is when one head position is driven by another and the gaze reenactment is where the direction of the eyes and the position of the eyelids are driven by another. Body reenactment uses human pose synthesis or pose transfer [45]. Everyone can dance now is an example of this type of deepfake[9].

**Video synthesis** is when one creates a video without a target to base it off of. Editing and synthesis are very similar in the regards that you are creating a new video when editing while synthesis you are creating an entire new video [45]. From video synthesis it was found that neural textures can be used to render/manipulate existing video content in static and dynamic environments in real time [66].

**FaceSwap:** The last category of video deepfakes is FaceSwap which is when someone's face in an image or video is replaced with another persons face([3]).

## 2.4 Image Deepfakes:

The last category discussed for deepfake technology is image deepfakes. Figure 5 provides the readers with a quick summary of image deepfake's generation and detection features and frameworks.

**Faceswap** One of the subcategories is faceswap. Snapchat was the beginning of the face swap technology available to the public [63]. Fakeapp is still the most popular face swapping app at the moment when it went viral around the world for showing people what they would look like when they were older and doing gender swaps[1].

**Synthesis** Image synthesis can allow someone to make a new AI generated image . It is easy to make deepfakes GANs now more than ever and there have been instances of synthesizing images that use GANs. NVIDIA's 112 can make endless variations of the same image [32]. StyleGAN2 is helping detect image deepfakes as it can see if the picture is generated by a network or not[32].

**Editing** Editing images using photoshop tools have been used for many years. However, image editing using AI tools has proposed robust way to edit the images vastly.

From the research conducted there seemed to be a plethora of papers written for image detection and creation as seen in the diagram 5. The survey performed by [69] has evaluated different image (focusing on face) manipulation as well as detection techniques.

## 2.5 Discussion and Future directions

This section firstly presents the critical discussion, analysis and summarization regarding the compiled works focusing on audio deepfake generation. Then, a summarization of the current techniques as well as future directions against deepfake is presented. Table 2 summarizes the key papers related to audio deepfake surveyed. Table 3 summarizes the key papers of the other types of deepfakes surveyed.

*2.5.1 Deepfake generation.* In deepfake generation, the most significant aspect is how believable it is to the victim, that means "deepfake quality".

**Data vs Quality (MOS)** The Mean Opinion Score (MOS) is "the arithmetical mean of individual ratings given by different users". MOS has been used in many researches surveyed here to identify the quality of the audio. Given our evaluation of different audio deepfake frameworks' performance, the Mean Opinion Score (MOS) of the generated audio is better when the framework is trained using single speaker datasets([52]; [55], [35], [37]. It means that training the models using multi-speaker datasets to generate natural audio samples could be challenging. MelNet [71] which has used a 140 hour single speaker dataset, as well as VoxCeleb2 [15] multi-speaker dataset, has a better performance than the previous works. The VoxCeleb2 dataset contains over 2,000 hours of audio data with real world noise. In addition, the dataset is captured from speakers of 145 different nationalities including different accents, ages and languages. The researchers are highly recommended to used different multi-speaker data such as VoxCeleb2 dataset and evaluate the obtained generalization.

**Sampling Frequency (kHz) vs Quality (MOS):** When the sampling frequency (sampling rate) of the audio deepfakes is less than 16kHz, perceived speech quality of audio deepfakes drops significantly, and the higher sampling rate may give way to higher audio quality ([57]). For future research the impact of different sampling rates on the audio deepfake quality could be investigated.

**Availability vs Quality** We also found that the more availability and reproducibility, the more development the technology will have. The frameworks including their code as well as the datasets used that are available publicly (e.g., [55, 60, 62, 71, 74]) are more likely to be used for nefarious purposes or research, so they will be more developed. It is recommended that academic centers prepare a researching environment to share deepfake related frameworks and datasets with just researchers.

Using Other deepfake types for a certain type: As we could see, a framework that has been proposed for generation of a certain type of deepfake, could be used for generation of another type of deepfake with some changes. For example, CycleGAN and StarGAN are two frameworks for image deepfake generation that are used as the base of two audio deepfake frameworks [17, 30], which can work with non-parallel data not just parallel ones. Data conditions for VC could be parallel or non-parallel. Parallel VC datasets refer to the datasets with utterances of the same linguistic content, but uttered by different people [80], but in practice, non-parallel VC which is more challenging is needed. It seems that more work should be done regarding audio deepfake frameworks using non-parallel data, and in this way researchers may use image deepfake frameworks as the base of their proposed framework.

*2.5.2 Future defense against deepfakes.* The approaches that are mentioned in the following paragraphs offer a modest defense against deepfakes.

**Prevention** For prevention of deepfakes some suggest blockchain and other distributed ledger technologies (DLTs) be used for finding data provenance and tracing the information ([10, 18, 33, 78]). Extracting and comparing affective cues corresponding to perceived emotions from the digital content is also proposed as a way to combat deepfakes [46]. Some recommend the content be ranked by participants and AI regarding if it is fake or real [11]. For future directions, deepfake prevention is the area that needs more attention. Especially, researchers could extend using DLTs for digital content traceability, as well as using affective computing to combat deepfakes.

**Mitigation** It might be better to keep some detection tools proprietary only to people who need it like fact checkers for reporters. This is so those making the generation models, perhaps for nefarious purposes, would not know exactly what features make it easier to detect a deepfake like. Additionally, the journals as well as academic centers can make researchers who work on extending deepfake generation frameworks, propose a strong method for detecting the deepfakes generated by their frameworks (e.g., [9] has proposed it for their framework "Everybody Dance Now").

**Detection** In the following sentences, we present our summarization and future directions about the spoofing detection systems focusing on **"audio deepfakes"**. In audio deepfake replay attack detection some of the frameworks have been proposed by working on the features which are fed into the network (e.g., [75]). Others have improved the networks used or have worked on both networks and features simultaneously (e.g., [24, 38, 39]). Another category of audio deepfake detection systems aims to detect speech synthesis as well as voice conversion. Most of them use different DNNs such as ResNet (e.g., [12, 13]) to detect spoofing. Additionally, some of the audio spoofing detection methods have been extended by working on the features which are fed into the network (e.g., [8]). While others have changed the networks used or have improved both networks and features (e.g., [6, 14, 44, 53, 59, 73]). Given the fact that one of the most important deepfake detection challenges is "generalization", researchers are highly recommended to work on generalization by changing or improving both of the networks and features as well as defining different loss functions (e.g., [12, 82]).

Given the aforementioned categories, we summarized some of the most important references regarding audio deepfakes which are used in this survey in table 2.

The references which are about the other deepfake types are summarized in Table 3. For text deepfakes, a very rich summarization is available [16, 22], therefore we only mentioned three new works in the text deepfake area below.

Additionally, for visual deepfakes (image and video) there are some more surveys (e.g., [45, 51, 81].)

## 2.6 Conclusion:

People not just in this research field but the everyday person, need to be aware of deepfakes and the harm they can cause to minimize the adverse effects. Also, we need to question what we see and hear online since the content can be misleading. The categories of deepfakes were broken down into four categories: audio, video, photo, and textual. There were also subcategories discussed in each of the main four categories along with the advantages, disadvantages, and summary of the methods for each subcategory. In addition, in this research, we have focused on audio deepfake generation and detection. We have provided a summary overview of how the technologies which are used to create or detect audio deepfakes work, and also the details of their architectures. We hope this survey serves as a guide for people who are interested in understanding and preventing malicious deepfakes, and those who want to use deepfakes for well-meaning purposes. More research needs to be done in the field of audio and text deepfakes, especially audio since there is already a plethora of detection for different categories of textual deepfakes, specifically in the category of fake news.

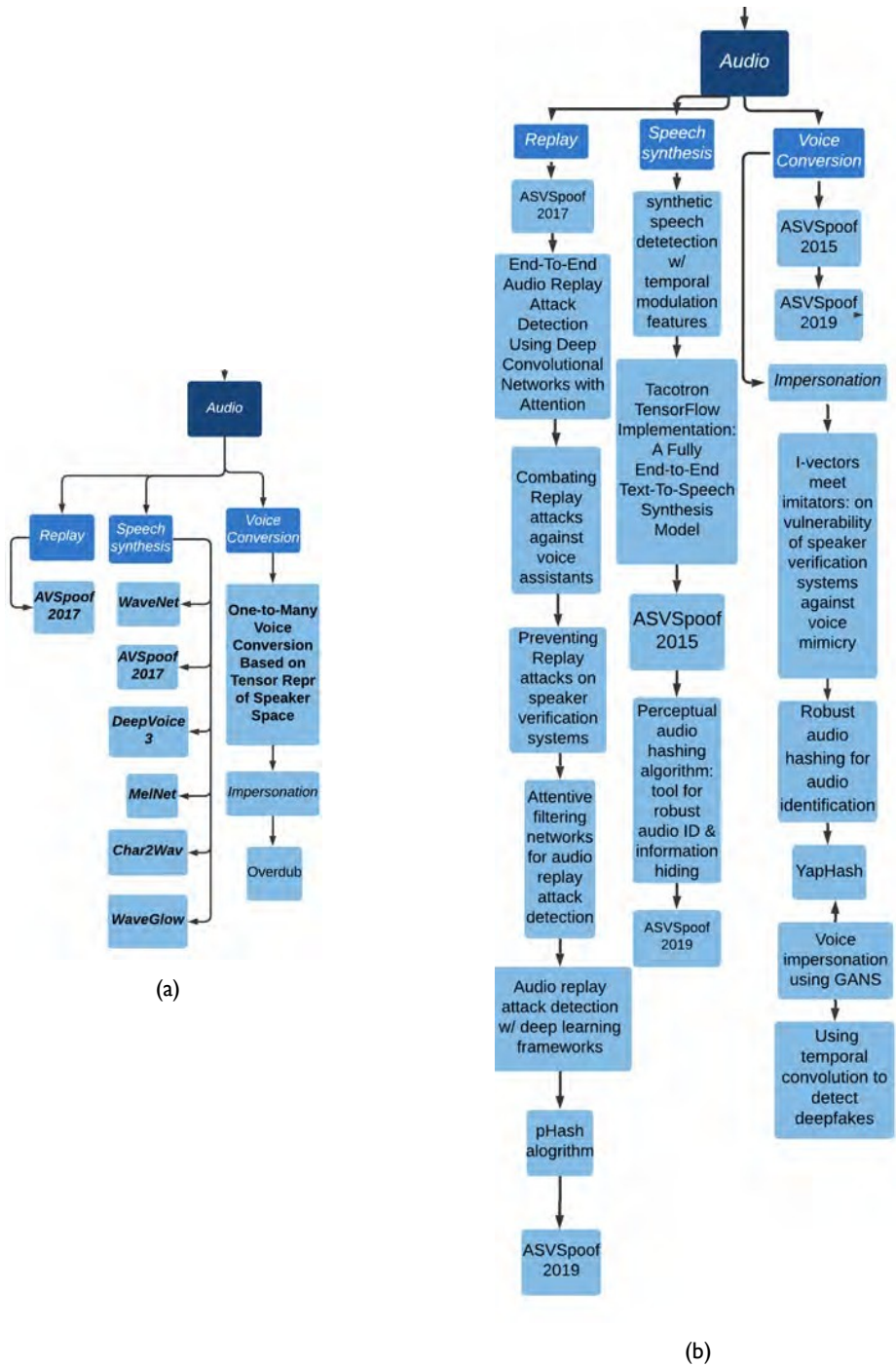

Fig. 2. a: Audio Deepfake Generation, b: Audio Deepfake Detection

Table 1. Quick guide for Audio Deepfake tools

| Type | Name | Ref | Sample | Key Features |
|---|---|---|---|---|
| Replay/ Detection | Replay attack end-to-end detection | [70] | https://mohitjaindr.github.io/pdfs/c20-interspeech-2018.pdf | Contains a visual attention mechanism on time-frequency representations of speeches that uses group delay features and ResNet-18 architecture. The model works perfectly with an Equal Error Rate of 0 percent |
| Synthesis/ Creation | Char2Wav | [62] | https://github.com/gcunhase/PaperNotes/blob/master/notes/. char2wav.md | Reader (frontend): Bidirectional RNN, transform the text into linguistic features" Neural vocoder (backend): Conditional SampleRNN: takes the linguistic features as input and creates the corresponding audio. |
| Synthesis/ Creation | Tacotron2 | [60] | https://github.com/ NVIDIA/tacotron2 | End to end Text to Speech model that uses recurrent sequence-to-sequence feature prediction network (for Embedding characters to melspectorgrams) and a modified WaveNet. |
| Synthesis/ Creation | VoCo | [29] | https:// www.youtube.com /watch?v=RB7upq8nz IU | Not open source. Includes text to speech and voice conversion of the text-based editing, pitch profile, manual editing of length and amplitude. |
| Synthesis/ Creation | WaveGlow | [57] | https://github.com /NVIDIA/waveglow | It combines insights from Glow and WaveNet to be able to provide fast, efficient and high-quality audio synthesis, without the need for auto-regression. It uses only a single network trained using only a single cost function: maximizing the likelihood of the training data, which makes the training procedure simple and stable |
| Synthesis/ Creation | Tacotron | [74] | https://github.com /Kyubyong/tacotron | End to end text to speech model, creates audio directly from text. |
| Synthesis/ Creation | MelNet | [71] | https://github.com/Deepest-Project/MelNet | It is introduced as a generative model for audio which can capture longer-range dependencies for the first time in the TTS area. MelNet couples a fine-grained autoregressive model and a multiscale generation procedure to jointly capture local and global structure. |
| Synthesis/ Creation | Deep Voice 3 | [55] | https://github.com/Kyubyong/deepvoice3 | A fully convolutional attention based neural for TTS that can create high-quality audio samples. |
| Synthesis/ Creation | Wavenet | [52] | https://github.com/ibab/tensorflow-wavenet | It uses Causal Convolutional layers and Dilated Causal Convolutional layers to create high quality audio deepfake. |

| Type | Name | Ref | Sample | Key Features |
|---|---|---|---|---|
| Synthesis/ Creation | GAN based Speech Synthesis | [58] | https://github.com /r9y9/gantts | Statistical parametric method for speech synthesis based on GANs |
| Synthesis/ Creation | HiFi-GAN | [35] | https://github.com /jik876/hifi-gan | A GAN based speech synthesis framework which outperformed a lot of the previous works. |
| Synthesis/ Creation | MelGAN | [37] | https://github.com /seungwonpark/melgan | non-auto-regressive, fast, fully convolutional with significantly fewer parameters than the other frameworks. |
| Voice Conversion/ Impersonation/ Creation | a GAN based model | [19] | Not Found | Transferring style from one speaker to another. Obtained from huge modifications on the DiscoGAN |
| Voice Conversion/ Impersonation/ Creation | CycleGAN-VC | [17] | https://github.com/ jackaduma/CycleGAN-VC2 | A VC system based on CycleGAN. A nonparallel VC method that only learns one-to-one-mappings |
| Voice Conversion/ Creation | StarGAN-VC | [30] | https://github.com/ liusongxiang/ StarGAN-Voice-Conversion | It has developed StarGAN (Choi et al., 2018) to a VC system that allows non-parallel many-to-many VC. There is a generator that takes an acoustic feature sequence instead of a single-frame acoustic feature as an input and outputs an acoustic feature sequence of the same length. (same as Kaneko et al. (2017) papers. |
| Voice Conversion/ Impersonation/ Creation | SINGAN | [61] | Not Found | GAN-based model for singing VC. |
| VC and SS/ Detection | - | [12] | Not found | Overcoming the generalization challenge by using: 1) large margin cosine loss function (LMCL) 2) online frequency masking augmentation that forces the neural network to learn more robust feature embeddings. |
| VC and SS/ Detection | - | [82] | https://github.com/ yzyouzhang/ AIR-ASVspoof | An attempt to detect unknown syn-thetic voice spoofing attacks using one-class learning. It compacts the bonafide speech representation and injects an angular margin to separate the spoofing attacks in the embedding space. |
| VC, SS and Replay attack/ Detection | - | [13] | Not found | Inspired by the success of ResNet in image recognition, they used it for automatic audio spoofing detection, and reduced the EER by 18 percent |

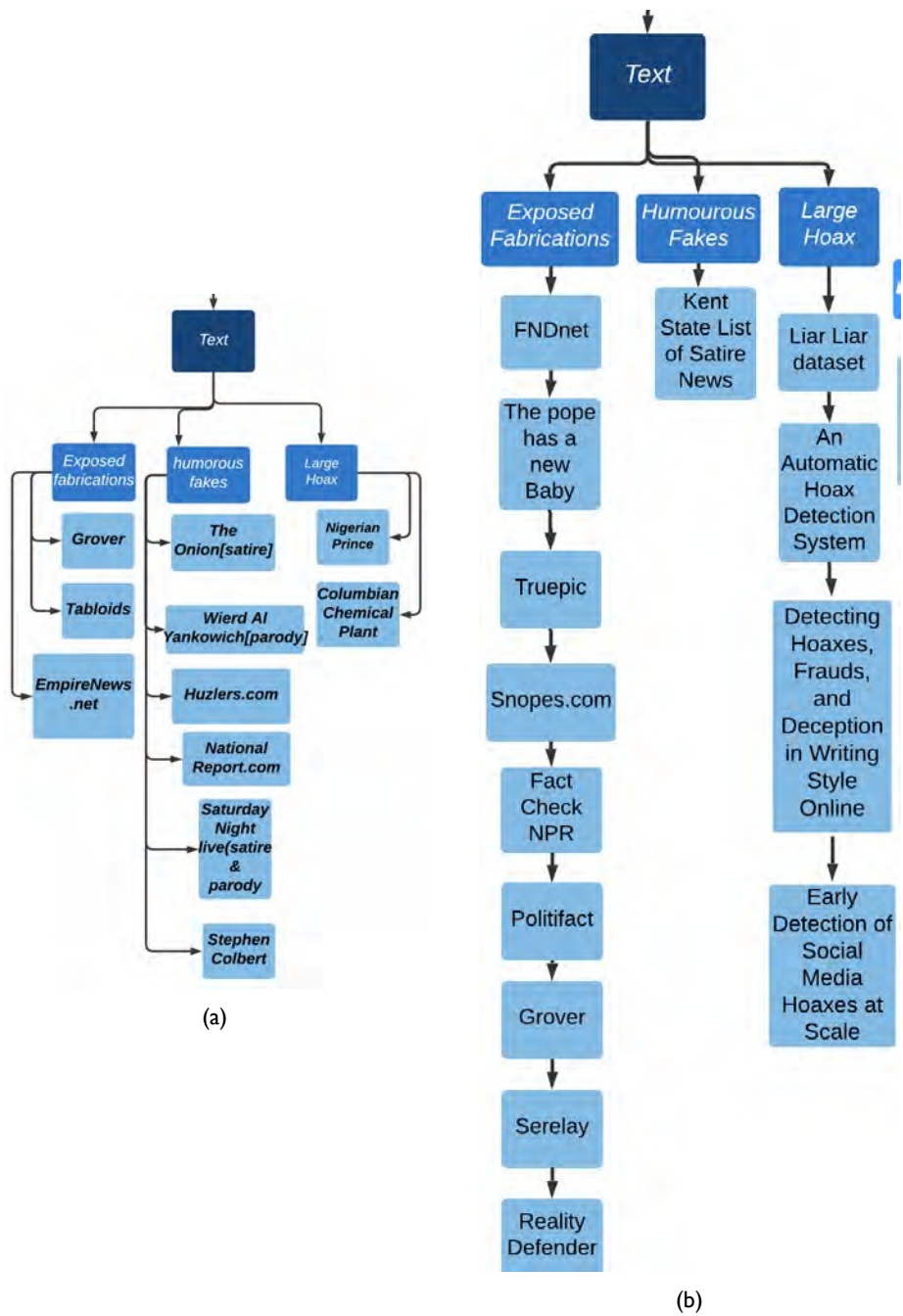

Fig. 3. a: Text Deepfake Generation, b: Text Deepfake Detection

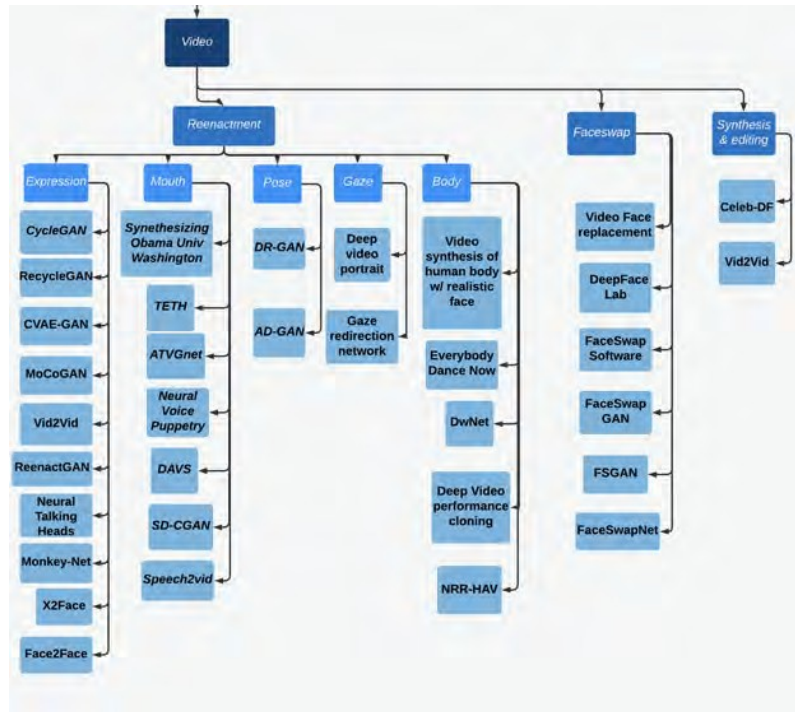

(a)

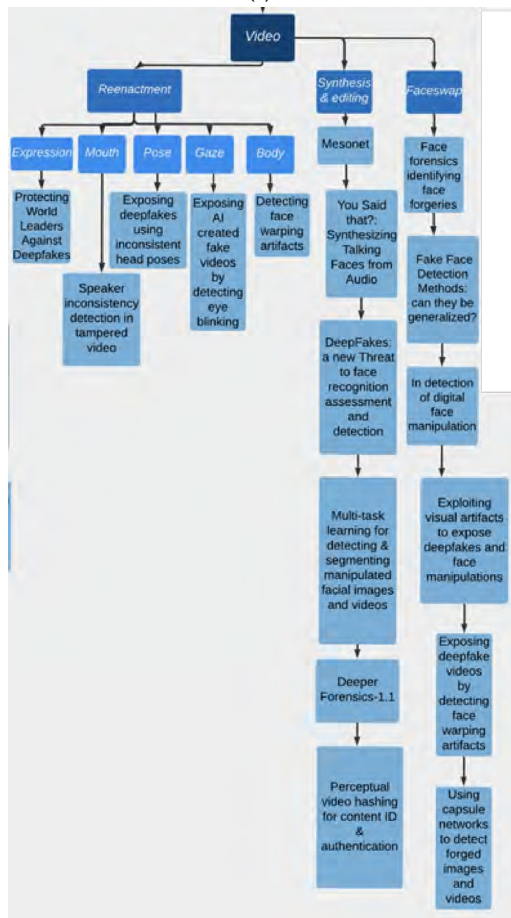

(b)

Fig. 4. a: Video Deepfake Generation, b: Video Deepfake Detection

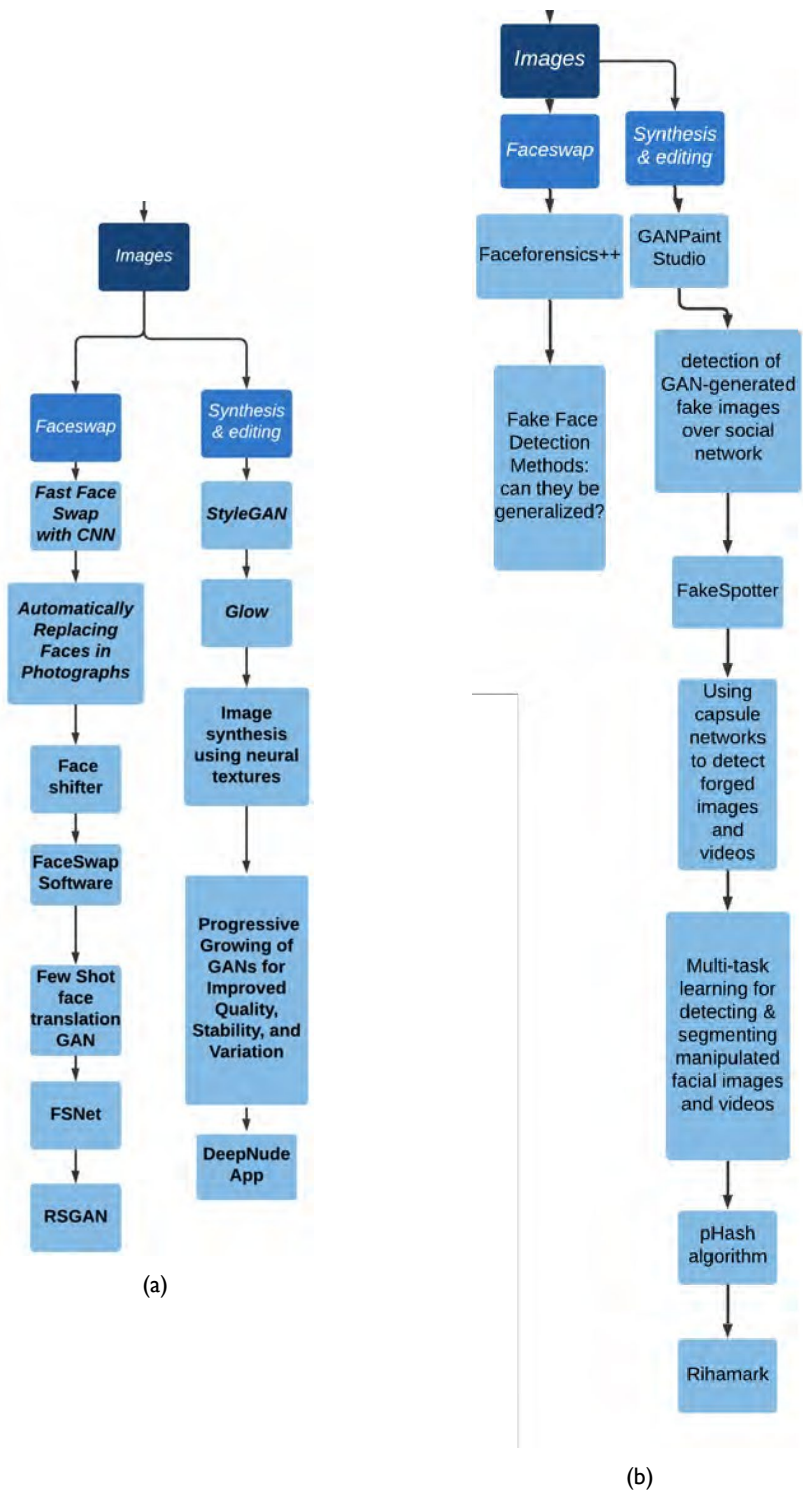

Fig. 5. a: Image Deepfake Generation, b: Image Deepfake Detection

Table 2. Summarization of the **Audio Deepfake** papers surveyed

| Name | Ref | Architecture | Dataset | Objectives | Metrics | Results |
|---|---|---|---|---|---|---|
| Replay attack end-to-end detection | [70] | Group delay, ResNet-18 (it also includes Global Average Pooling, Fully Connected layers) | ASVspoof 2017 [34] | An end-to-end deep learning framework for audio replay attack detection. | EER | Outperformed previous related works with Equal Error rate= 0 percent |
| Wavenet | [52] | Causal Convolutional layers, Dilated Causal Convolutional layers | To measure WaveNet Audio performance: For Multi Speaker Speech Generation: VCTK is used, and the dataset contained 44 hours for 109 speakers. For TTS: The same single-speaker speech databases from which Google's North American English and Mandarin Chinese TTS systems are built. For music audio modelling: the MagnaTagATune and the YouTube piano datasets. | High quality audio deepfake generation in three areas: Speaker Speech Generation, TTS and music audio modelling. | MOS scale for quality | Generation of raw speech signals with subjective naturalness never before reported in the TTS area. A new architecture based on dilated causal convolutions are developed. When conditioned on a speaker identity, a single model can generate different voices using WaveNet. The architecture of WaveNet gives great results when tested on a small speech recognition dataset. It also is so good in generating other audio modalities such as music. |
| VoCo | [29] | Forced alignment algorithm for corpus preparation TTS selection and preparation using MCD metric Building a voice convertor Synthesis and blending | CMU Arctic dataset | Editing audios using texts. for example to insert a new word for emphasis or replace a misspoken word. | Mel-Cepstral Distortion (MCD) Euclidean distance, MOS scale for quality | A system is presented that can synthesize a new word or short phrase, replace or insert it in the context of the existing speech |

| Name | Ref | Architecture | Dataset | Objectives | Metrics | Results |
|---|---|---|---|---|---|---|
| Tacotron | [74] | PreNet, Attention RNN, Decoder RNN, CBHG (1-D convolution bank + highway network + bidirectional GRU), Griffin-Lim reconstruction | an Internal North American English | An end-to-end generative TTS model that synthesizes speech directly from text (characters) | Visual Comparisons, MOS scale for quality | An end-to-end generative TTS model that achieved 3.82 subjective 5-scale score on US English. |
| Tacotron 2 | [60] | PreNet, three convolutional layers, bidirectional LSTM, attention, 2 LSTM layers (recurrent sequence-to-sequence feature prediction network), Linear Projection, Modified WaveNet as vocode | an Internal US English Dataset | A NN architecture for speech synthesis directly from text. | MOS scale for quality | MOS = 4.53 comparable with MOS= 4.58 that is assigned to a professionally recorded speech. |
| WaveGlow | [57] | It uses only a single network and a single cost function: maximizing the likelihood of the training data | Amazon Mechanical Turk | Presenting a "flow-based network capable of generating high quality speech from mel-spectrograms". | MOS scale for quality | It produces audio samples at a rate of more than 500 kHz on an NVIDIA V100 GPU. The MOS of it shows that its generated audio is as good as the best public WaveNet audio sample. |
| Char2Wav | [62] | The Reader: (bidirectional RNN as the encoder, attention-based RNN as decoder) followed by the vocoder ( neural vocoder and sampleRNN). | For being conditioned on English phonemes and texts is VCTK. For Spanish text is DIMEX-100 | An end-to-end model for speech synthesis | No metric. Just some samples are given | Not contain any comprehensive quantitative analysis of the results, but shows some samples and their corresponding alignments to the texts. |

| Name | Ref | Architecture | Dataset | Objectives | Metrics | Results |
|---|---|---|---|---|---|---|
| Melnet | [71] | Entirely recurrent architecture. It has three stacks, respectively: time delayed (Multiple layers of multidimensional RNNs), the optional centralized stack (an RNN), the frequency-delayed stack: (a one-dimensional RNN) | For unconditional generation: (Blizzard, MAESTRO, VoxCeleb2) For TTS: (TED-LUM 3 and Blizzard) | To have a generative network that can capture high-level structure that emerges on the scale of several seconds (subsampled spectrogram) in addition to details and high resolution (an iterative upsampling procedure) | Which model generates samples with longer-term structure answered by some human evaluators. | Instead of 1D time domain waveforms, it modelled 2D time-frequency representations such as spectrograms. It enabled fully end-to-end TTS that can generate audio with longer-term structure. |
| Deep Voice 3 | [55] | Encoder: (PreNet, Convolutional Blocks, PostNet), Decoder: (in an autoregressive manner: PreNet, Attention Blocks and Causal Convolutional layers, Fully Connected Layers) then it followed by the convertor and chosen vocoder. | Single-speaker synthesis: an internal US English dataset. Multi-speaker synthesis: VCTK and LibriSpeech [54] | A new high-quality framework (A fully convolutional attention based neural) for TTS. The common error modes of attention-based speech synthesis networks and ways to mitigate them. Comparing some waveform synthesis models with each other. | Training iteration and achieving convergence time, MOS scale for quality | It is compared with Tacotron and is really faster in training iteration time and achieving convergence. The achieved range of MOS for single-speaker synthesis: (3.62-3.78) The achieved range of MOS for multiple-speaker synthesis for VCTK: (3.01-3.44). and for LibriSpeech (2.37-2.89) |
| HiFi-GAN | [35] | Generator (CNN) and two D Discriminators: multi-scale and multi-period. Two losses are added during the training. A Multi-Receptive Field Fusion module is added to the G | LJSpeech dataset [26] | Proposing a framework for efficient and high fidelity speech synthesis based on GAN | MOS scale for quality | It outperformed WaveGlow in the end-to-end setting and some other frameworks, but the quality of the ground truth is still better. |

| Name | Ref | Architecture | Dataset | Objectives | Metrics | Results |
|---|---|---|---|---|---|---|
| MelGAN | [37] | GAN-based: G: (Convolutional layer, Upsampling layers, Residual Stacks with dilated convolutional block). D: (Each discriminator block contains Downsampling layer as well as some convolutional layers) | LJSpeech dataset [26] | Using GAN for producing high quality coherent waveforms) | Number of parameters, Speed (in kHz), MOS | Its pytorch implementation ran more than 2 times faster in real time on a CPU and 100x faster than realtime on a GTX 1080 Ti GPU, with no hardware specific optimization trick. It is comparable in quality to state-of-the-art high capacity WaveNet-based models, but not better than them. |
| CycleGAN-VC | [17] | Extracting two types of features of speech: The mel-cepstrum, (fundamental frequency) and aperiodicity bands. They are converted separately. GAN and linear conversion. | ALAGIN Japanese Speech Database | A nonparallel VC method | MOS scale for quality and similarity | Outperformed Merlin-based baseline (parallel VC) significantly. Slightly better than the state-of-the-art GAN-based parallel VC method |
| StartGAN-VC | [30] | Extracting two types of features of speech: The mel-cepstrum, (fundamental frequency) and aperiodicity bands. They are converted separately. GAN and linear conversion | ALAGIN Japanese Speech Database | A nonparallel VC method | MOS scale for quality and similarity | Outperformed Merlin-based baseline (parallel VC) significantly. Slightly better than the state-of-the-art GAN-based parallel VC method. |
| SINGAN | [61] | In the training phase WORLD vocoder is used. The run-time phase includes WORLD vocoder, GAN and another WORLD vocoder. | NUS Sung and Spoken LyricsCorpus (NUS-48E corpus) (Duan et al., 2013) | Using GAN for SVC (Singing Voice Conversion) | MOS, Preference score | Outperformed the DNN-based traditional SVC model. |

| Name | Ref | Architecture | Dataset | Objectives | Metrics | Results |
|---|---|---|---|---|---|---|
| GAN for Impersonation | [19] | GANs for style transfer, GANs for voice mimicry. The generative network: (6-layer CNN encoder and transposed 6-layer CNN). The discriminative network (7 layer CNN with adaptive pooling.) | TIDIGITS | Presenting GAN-based model for voice impersonation | The signal-to-noise (SNR) ratio test using the standard NIST STNR method and the WADA SNR method | The WADA test results are around 100 db. The STNR shows the generated data has good quality. |
| A spoofing detection system (ResNet-based) | [12] | The input of the ResNet is 60-dimensional linear filter banks (LFBs) that are extracted from raw audio. In the training phase FreqAugmentlayer and large margin cosine loss are used; same training utterances are fed into ResNet to extract spoofing embeddings. They are utilized to train the backend genuine-vs-spoof classifier. | 1) ASVspoof 2019 logical access (LA)dataset. 2) A noisy version of the ASVspoof 2019 dataset. 3) A copy of the dataset that is logically replayed through the telephony channel. | Proposing a spoofing detection system that overcomes the generalization challenge | EER | Reduced ERR significantly (from 4.04 percent to 1.26 percent) |
| A spoofing detection system (ResNet-based) | [13] | Features: Constant Q Cepstral Coefficients and Mel Frequency Cepstral Coefficients Classifiers: Gaussian Mixture Models, DNN and ResNet | ASVspoof2017 [34] dataset | Using ResNet for automatic audio spoofing detection | EER | Outperformed the best single-model system by reducing EER 18 percent relatively. |

| Name | Ref | Architecture | Dataset | Objectives | Metrics | Results |
|---|---|---|---|---|---|---|
| One-Class Learning Towards Synthetic Voice-Spoofing Detection | [82] | It has proposed the OC-Softmax loss function to solve the generalization problem | The development and evaluation sets of ASVspoof2019 Challenge logical access scenario | To detect unseen voice spoofing attacks using one-class learning, and solve the generalization problem of the previous detection methods | EER, countermeasure (CM) score | It has achieved an EER of 2.19 percent, and outperformed all existing single systems (i.e., those without model ensemble) |

Table 3. Summarization of the works surveyed regarding the other Deepfake types

| Text Deepfake | | | | | | |
|---|---|---|---|---|---|---|
| Name | Ref | Architecture | Dataset | Objectives | Metrics | Results |
| DGSAN | [47] | The framework contains iterations that each new generator is defined based on the last discriminator. For each iteration, while not converged, the framework samples from real data and generates fake data. It considers the generator found in the last step is called Qold and tries to generate Qnew from it. It also uses a discriminator between the real distribution and Qold, and tried to make Qnew optimal | COCO Image Captions, EMNLP2017 WMT News, Chinese Poems | Solving the problem of the existing GAN based methods with generating discrete data. | NLL, BLEU, Self-BLEU, MS-Jaccard | A new GAN-based framework to generate discrete data in which there is no need to pass the gradient to the generator. It has really better performance in the MS-Jaccard metric. |

| Name | Ref | Architecture | Dataset | Objectives | Metrics | Results |
|------|-----|--------------|---------|------------|---------|---------|
| TextGAIL | [76] | A generative adversarial imitation learning framework for text generation is presented that utilizes huge pre-trained language models (pre-trained GPT-2 and RoBERTa) to provide a reliable guiding signal in the discriminators of the GANs. A contrastive discriminator, and proximal policy optimization (PPO) is applied to improve text generation. | COCO Image Captions, EMNLP2017 WMT News | Providing a reliable guiding signal in the discriminators of the GANs for text generation. Improving text generation performance using GANs. | NLL, BLEU, Self-BLEU, Perplexity | Unlike most of the previous GAN text generation frameworks, TextGAIL obtained better performance in terms of both quality and diversity than the MLE baseline |
| FGGAN | [77] | A GAN-based model, but it utilizes a feature guidance module for text features extraction from the discriminator network. These text features are converted into feature guidance vectors which are fed into the generator network to enhance guiding signal. Text semantic rules are also formulated. | COCO Image Captions, Synthetic Data, Chinese Poems | Solving the problem of weak feedback (guiding signal) from the discriminator network to improve GAN-based text generation performance | NLL, BLEU | Better than existing baselines in terms of sentence quality |

| Video Deepfake | | | | | | |
|---|---|---|---|---|---|---|
| **Name** | **Ref** | **Architecture** | **Dataset** | **Objectives** | **Metrics** | **Results** |
| Everybody Dance Now | [9] | TPose Encoding: pretrained Open-Pose + global pose normalization. Pose to vid Translation: the updated GANs (for temporal smoothing and Face GAN) Optimizing GANs with the objective | 62 subject set of short 1920 × 1080 resolution dancing videos from YouTube | a simple method for "do as I do" motion transfer | SSIM.: Structural Similarity. LPIPS: Learned Perceptual Image Patch Similarity. Pose distance is evaluated between input and target. | High quality deepfake videos |
| - | [79] | A Meta learning: Embeddor, a Generative network and discriminative network (GAN) | VoxCeleb1 and Voxceleb2 | Creating personalized talking head models using just a few images of a person or even one image. | SSIM, CSIM, USER | It presented a framework with so few-shot capability that is considered a meta-learning GAN model. It has been trained on a huge dataset of video. Then, it can learn one or few shot learning of neural talking head models of unseen people. |
| - | [65] | RNNs are used. A RNN learns the mapping from raw audio features to mouth shapes. Given the mouth shape at each time-instant, mouth texture is synthesized, then it is composed with proper 3D pose matching after retiming. | For training 300 weekly addresses spanning 2009 to 2016 are downloaded. Each address lasts about 3 minutes on average, therefore totally 17 hours of video are used | Synthesizing video from audio in the region around the mouth | More natural | High quality video deepfake of President Barack Obama speaking with accurate lip sync |

| Name | Ref | Architecture | Dataset | Objectives | Metrics | Results |
|---|---|---|---|---|---|---|
| Deferred Neural Rendering | [66] | The rendering network is based on a U-Net [25]. Encoder:(some convolutional layers each with instance normalization and a ReLU activation). Decoder:(mirrors the encoder). A TanH activation as inPix2Pix [25] is used for the final output layer. | Synthetic sequence including 1000 random training views and a smooth trajectory of 1000 different test views on the hemisphere. For facial reenactment: They used 650 training images for Macron, 2400 for Obama, and 2400 for Sequence 17. | A system that combines the traditional graphic pipeline with learnable components of ML to use of imperfect 3D content for producing photo-realistic (re-)rendering | MSE | Neural Rendering is introduced for photo-realistic image synthesis based on imperfect 3D contents at real-time rates. Neural Textures are presented for novel view synthesis in static scenes and for editing dynamic objects. It is faster than regular reenactment, asking only a few milliseconds input for high-resolution output. |
| - | [50] | AUnsupervised learning and progressive training of multi-subject face swapping: Normalization all available examples to 1024X1024 resolution. 1) Embedding images using a common encoder, mapping back them to the pixel space using a desired decoder (a multi-way decoder allows for generating different outputs). 2) Face alignment and landmark stability 3) Contrast-Preserving, Multi-Band Compositing | They created their own dataset not publicly available. | High resolution face-swapping pipeline at megapixel resolution. | No special metric, but the results are more realistic in comparison to the other state-of-the art face-swapping mode. | The first method capable of rendering photo-realistic and temporally coherent results at megapixel resolution. The importance of progressive training for high-resolution face-swapping is proved. Providing a landmark stabilization procedure that mitigates the temporal instabilities in the high-resolution domain. It has been compared with three open-source approaches that were considered as the state of the art in facial appearance transfer |

| Image Deepfake | | | | | | |
|---|---|---|---|---|---|---|
| **Name** | **Ref** | **Architecture** | **Dataset** | **Objectives** | **Metrics** | **Results** |
| NVIDIA's Style-GAN2 | [32] | The style of the blocks includes: Modulation followed by Convolutional layers (3X3), then normalization. Several changes are done on the original StyleGAN [31] to obtain the revised architecture. For example: the addition of biases and the noise broadcast operation are moved to the outside active area of a style. The revised architecture makes it possible to replace instance normalization with a "demodulation" operation. | FFHQ and LSUN CAR | Revising and improving the StyleGAN framework [31]. | FID, PPL, Precision, Recall | The proposed framework (StyleGAN which was state-of-the-art in data-driven unconditional generative image modeling) is improved and analyzed in terms of existing distribution quality metrics as well as perceived image quality. |
| - | [41] | GAN: Generative network: based on Markov random field (MRF) models. Discriminative network: trained deep convolutional neural networks (dCNNs). GitHub Code: https://github.com/chuanli11/CNNMRF | - | Synthesizing 2D images | No special metric, but the results are better than previous works especially in art-work synthesis | The method can transfer both photorealistic and non-photorealistic styles to a new image. The combination of the discriminative power of a deep neural network with classical MRFs based models gives high-quality image synthesis. |